\documentclass{ws-procs9x6}

 \newcommand{\zr}[1]{\mbox{\hspace*{#1em}}}

 \newcommand{\ZZ}{\mbox{\sf Z\zr{-0.45}Z}}
\def\ltap{\;\raisebox{-.4ex}{\rlap{$\sim$}} \raisebox{.4ex}{$<$}\;}  

\renewcommand\bibname{}%
\def\headnote{\hfill\hbox to \trimwidth{\footnotesize 
\hspace*{.7cm} Presented at {\it
Confinement V}, Gargnano, Italy, September
 10-14, 2002  \hfill FAU-TP3-02/30\hfill}}%

\let\trimmarks\mymarks

\begin{document}

\title{\vspace*{-.7cm}
More on Electric and Magnetic Fluxes in $SU(2)$\\
\vspace{-.5cm}}

\author{Lorenz von Smekal,}

\address{Institut f\"ur Theoretische Physik III,
Universit\"at Erlangen-N\"urnberg,\\
D-91058 Erlangen, Germany}

\author{\vspace{-.3cm}
{\small with} Philippe~de~Forcrand {\small and} Oliver Jahn}

\address{ETH-Z\"urich, CH-8093 Z\"urich and CERN Theory Division,\\
CH-1211 Geneva 23, Switzerland}  

\maketitle

\abstracts{\vspace*{-.4cm}
The free energies of static charges and center monopoles are given by their 
fluxes. While electric fluxes show the universal behaviour of the
deconfinement transition, the monopole free energies vanish in the
thermodynamic limit at all temperatures and are thus irrelevant for the
transition. Magnetic fluxes may, however, be used to measure the topological
susceptibility without cooling. \vspace*{-.3cm}}

In the pure $SU(N)$ gauge theory without quarks 't~Hooft's gauge-invari\-ant 
electric and magnetic fluxes describe, respectively, the effect of a static
fundamental color charge and a center mono\-pole in a finite
volume.\cite{tHo79} The partition function of a certain amount of
electric/magnetic flux yields the free energy of a static electric/magnetic
charge with boundary conditions to imitate the presence of its 'mirror'
(anti)charge in a neighboring box along the direction of the flux. 
To measure the free energies of fluxes one imposes, in a first step, 
't~Hooft's twisted b.c.'s to fix the total numbers modulo $N$ of
$\ZZ_N$-vortices through the 6 planes of the 4-dimensional Euclidean $1/T\!
\times\! L^3$ box.~In $SU(2)$ for example, twist in one plane corresponds to
an ensemble with an odd number of $\ZZ_2$-vortices through that plane. 
It differs by at least one from the periodic ensemble  
with an even number; and their free-energy difference is what it costs to add
one such vortex to the system. 

Qualitatively, the vortices lower their free energy by spreading in the plane
of the twist. At $T>0$ we thus distinguish between two types: 

{\bf Temporal twist} in a $1/T\!\times\! L$ plane is classified by a vector
$\vec k \!\in\! \ZZ_N^3$ parallel to the $L$-edge. With increasing
temperature $T$, the vortices are squeezed in such a plane more and
more. They can no-longer spread arbitrarily  and this is what drives the phase
transition. In the thermodynamic limit, their free energy approaches zero
(infinity) for $T$ below (above) $T_c$.\cite{deF02,Sme02}

{\bf Magnetic twist} is defined in a purely spatial plane.~It fixes the
conserved, $\ZZ_N$-valued and gauge-invariant magnetic flux $\vec m $
through that plane. Since the vortex can spread 
in such a plane independent of $T$, its free energy, or that of a static
center monopole, vanishes for $L\!\to\!\infty$ at all $T$.\cite{Sme02b}

The partition functions of fixed units of electric and 
magnetic fluxes, $\vec e, \vec m \!\in\! \ZZ_N^3$, which we denote 
by $Z_e(\vec e, \vec m)$, are obtained as 3-dimensional $\ZZ_N$-Fourier
transforms,  w.r.t. the temporal $\vec k$-twist, of those with twisted b.c.'s,
$Z_k(\vec k,\vec m)$.~Purely electric flux
yields the free energy of a static fundamental charge $\propto -\ln
Z_e(\vec e,0)$ in a well-defined (UV-regular) way.\cite{deF01}
Measuring in $SU(2)$ the ratios of different $Z_k$'s 
($Z_e$'s) for the various $\vec k$-twists (electric fluxes)
we demonstrated their Kramers-Wannier duality.\cite{deF02}   
This duality is the analogue in $SU(2)$ of that
between the Wilson loops of the 3d $\ZZ_2$-gauge theory and the 3d-Ising
spins reflecting the different realizations of the 3-dim.~{\em electric center
symmetry} in both phases.  

There is no analogue of magnetic flux in the 3d-spin systems.
Changing \linebreak the spatial $\vec m$-twist is a gauge-invariant way of
introducing one more static center monopole.~However, the monopole free
energy is exponentially 
suppressed with the spatial string tension $\sigma_s$ and thus tends to zero,
in  the thermodynamic limit, at all temperatures. At $T_c$, for example, 
our data yields $\sigma_s \!=\! (2.2 \pm 0.2) T_c^2$ 
consistent with the zero temperature $SU(2)$ string tension.\cite{Sme02b} 
We find $Z_k(\vec k,\vec m)\! \to\! Z_k(\vec k,0)$
for all twists, and $ Z_e(\vec e,\vec m)\! \to \! Z_e(\vec e,0)$
for all fluxes, with $L\!\to\!\infty$ at any $T$.
Magnetic fluxes are thus irrelevant for the phase transition.
The corresponding 3-dim.~{\em magnetic center symmetry} remains unbroken,
and center monopoles always 'condense'.

Nevertheless, magnetic twist can be used to 
fix the fractional content of the topological charge $Q=(\nu +\vec
k\!\cdot\!\vec m/N)$ (with $\nu\in\ZZ$). 
In $SU(2)$ it is half-odd integer for $\vec k\!\cdot\!\vec m = 1$ mod 2 and
integer otherwise. This may be used to extract the topological
susceptibility from differences, in the finite volume, between the integer
and half-odd integer sequences of topological sectors. Recall that with
twisted b.c.'s the $\theta$-sectors can be represented as
\begin{equation}
\textstyle 
Z_\theta(\vec k, \vec m,\theta ) = \exp\{-F_\theta (\vec k, \vec m,\theta
)/T\}  = \sum_\nu e^{-i\theta(\nu+\frac{\vec
k\vec m}{N})} \, Z_\nu(\vec k, \vec m,\nu ) \; , 
\end{equation}
from which for the topological susceptibility $\chi$, one obtains,
\begin{equation}
\chi\, L^3 = \frac{d^2}{d\theta^2} \, F_\theta (\vec k, \vec m,\theta
)\Big|_{\theta=0} =  T \, \frac{\sum_\nu (\nu+\frac{\vec
k\vec m}{N})^2 \, Z_\nu(\vec k, \vec m,\nu )}{ Z_\theta(\vec k, \vec m,0) }
\; .
\end{equation}
We expect $\!\chi\!$ to result from a density of (more or less) localised
objects. Local observables should, however, be 
insensitive to b.c.'s, at least to magnetic
twist which we may introduce without cost for sufficiently large $L$.  
We therefore assume that the partition functions factorize
$Z_k(\vec k, \vec m) \!\equiv\! Z_\theta(\vec k, \vec m,0) \!=\! \sum_\nu
w(Q) \widetilde Z(\vec k, \vec m) $, with a form of $w(Q)$ that does not depend
on the b.c.'s and a $Q$-independent factor $\widetilde Z $ that does.~The 
ratios with odd/even  $\vec k\!\cdot\vec m$, shown for various $L$
in Fig.~1, are fairly independent of the magnitude of
the twists thus confirming this factorization which predicts for them the
unique form $\sum_\nu w(\nu + 1/2) / \sum_\nu w(\nu )$.
Inserting a Gaussian topological charge distribution
$w(Q) = \exp\{ -Q^2/(2\Delta)\}$, as a first approximation to a more realistic
one,   
we obtain the fits shown in Fig.~1 together with
the rough estimate, $\chi = \langle Q^2\rangle/(L^3/T) \approx \Delta/(L^3/T) 
\approx (156(5)\mbox{MeV})^4 $, for $\Delta \gg 1/4$. 
Since $\Delta\!\propto\!L^3/T$, the last condition fails for smaller volumes
(larger $T$). Then, 
$\langle Q^2\rangle \equiv \sum_\nu Q^2 w(Q)/\sum_\nu w(Q) \not=\Delta $
(but $\langle Q^2\rangle/\Delta \!\to\! 0$ as can be verified by Poisson
resummation).  
This limits the applicability of the estimate to  $T \ltap
\{ 0.8, 1.1, 1.3\} T_c$ for the $LT = \{4/4, 6/4, 8/4\}$ lattices.

\begin{figure}[t]
\vspace*{-.5cm}
\centerline{\epsfxsize=3.8in\epsfbox{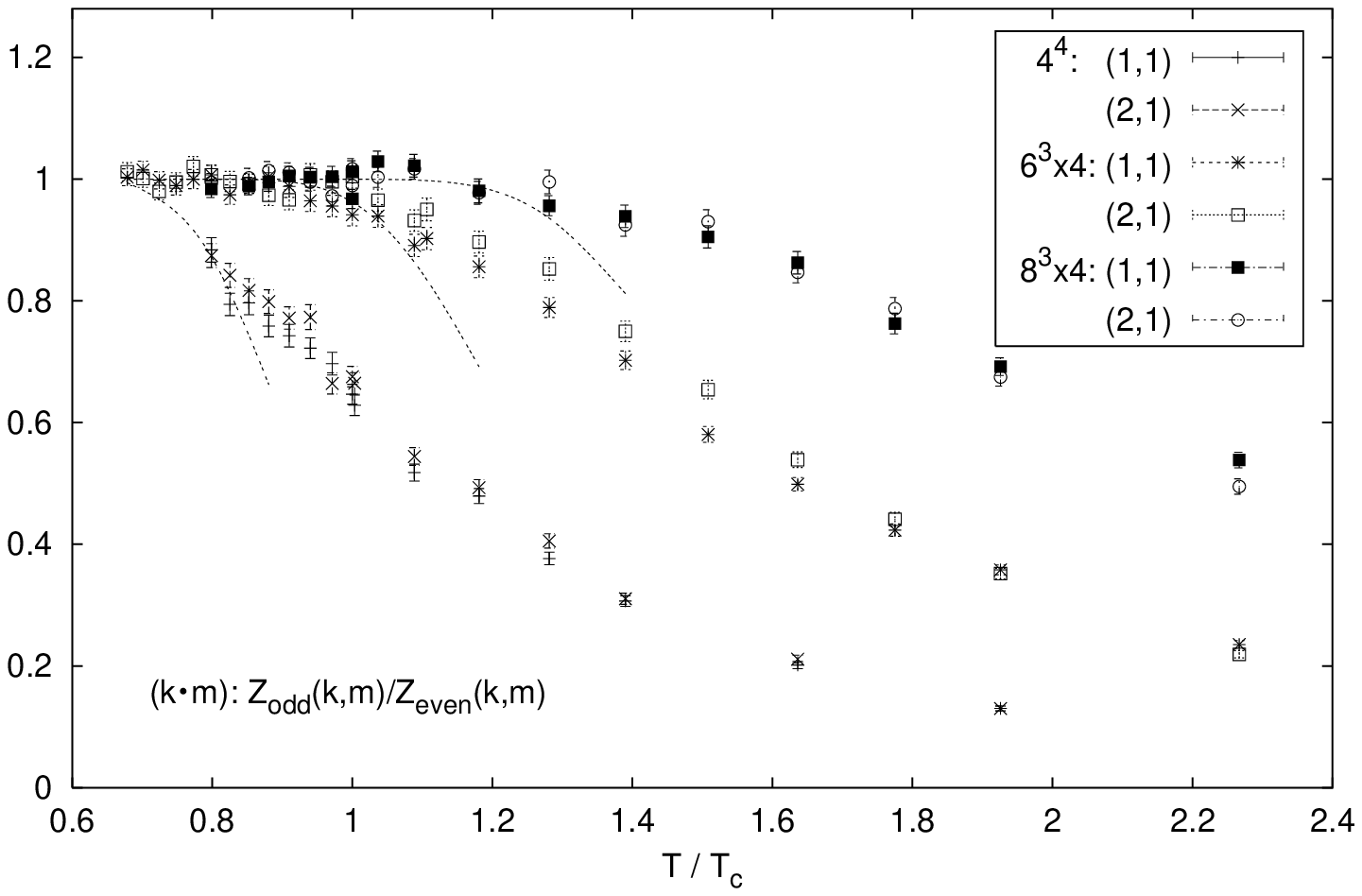}}   
\vspace*{-.5cm}
\caption{Ratios of partition functions of half-odd integer/integer
topological sectors.
\label{Fig1}}
\vspace*{-.35cm}
\end{figure}

Larger spatial lattice sizes are required to extend
the temperature range of the procedure upwards and to reliably extract a
temperature dependent topological susceptibility in a range around $T_c$. 
Encouragingly, our estimate for $\chi $ is of the right order, 
it extends to $T>T_c$, and our preliminary analysis suggests that
the topological susceptibility can be extracted without cooling, at least in
principle, by comparing via twisted b.c.'s sectors that differ only by the
fractional part of their topological charges.   

\smallskip

{\small\noindent
 Partially supported by grant \uppercase{SM} 70/1-1 of the 
\uppercase{D}eutsche
\uppercase{F}orschungsgemeinschaft.\\[-3pt] 
 Simulations were performed on the SGI Origin systems at the RRZE,
 Erlangen,\\[-3pt]
 and the ZHR, Dresden. L.v.S. thanks the organizers for this very nice
 conference.}

\end{document}